# Gaps in Representations of Hydropower Generation in Steady-State and Dynamic Models

Bhaskar Mitra, *Senior Member IEEE,* Sohom Datta, *Senior Member, IEEE,* Slaven Kincic, *Senior Member, IEEE*, Nader Samaan, *Senior Member, IEEE*, Abhishek Somani

*Abstract* — In the evolving power system, where new renewable resources continually displace conventional generation, conventional hydropower resources can be an important asset that helps to maintain reliability and flexibility. Varying climatic patterns do affect the operational pattern of hydropower. This would potentially play a vital role in meeting and delivering energy and meeting climate policy needs. Hydropower is one of the oldest forms of renewable energy resources, however, its dependency on water availability and other constraints are not well represented in power system steady state and dynamic models. This leads to multiple gaps in operations planning especially due to high intermittent renewable generation. Operating constraints and lack of high-quality data often become a barrier to hydropower modeling which leads to inconsistencies in reliability and operational planning studies resulting in unintentional blackouts or unforeseen situations. This paper identifies some of the gaps in hydro-based generation representation in steady-state and dynamic models and provides recommendations for their mitigation.

## I. INTRODUCTION

Continuous increases in penetration of renewables and retirement of traditional coal-fired and nuclear-based generators impose more burdens on the natural gas and hydro resources needed to provide regulation and load following. Hydropower provides opportunities of decarbonization through grid enhancement and grid resiliency. Development of pumped storage hydropower, run-of-river and coupling hydropower with storage helps the grid to become more resilient and better able to manage fluctuations in energy supply and demand. Hydropower generation involves understanding challenges posed by ecological conditions, soil and sediment behavior, stake holder considerations etc. However, when it comes to power system modeling understanding the challenges posed by dynamic modeling gaps, uncertainty quantification market policy considerations are not well understood. Advancements in modeling techniques, data availability, and computational resources will enable more accurate representation and optimization of hydropower generation within power systems, ultimately supporting the transition towards sustainable and resilient energy systems [1]. Between 2010 and the first quarter of 2019, U.S. power companies announced the retirement of more than 546 coal-fired power units, totaling about 102 GW of generating capacity. Plant owners intend to retire another 17 GW of coal-fired capacity by 2025, according to the U.S. Energy Information Administration (EIA) [2]. In the Western Interconnection, natural gas-based generation plays a major role; however, California also keeps downsizing its gas generation fleet. In 2018, California saw three big plant retirements: Encina (854 MW), Mandalay (560 MW), and Etiwanda (640 MW). This continues California's downward trend in natural gas power plant capacity. California's gas fleet peaked in 2013 with just over 47,000 MW of capacity, but California has shed roughly 5,000 MW since then. In the next 10 years there will be many more natural gas and coal plants retired within the Western Interconnection, and therefore more burdens placed on hydro-based resources [3]. As the power system continues to be flooded with intermittent resources, it becomes more important to accurately assess the role of hydro and its impact on the power grid [4]. While hydropower generation has been studied for decades, dependency of power generation on water availability and constraints in hydro operation are not well represented in power system models used in planning and operation of large-scale interconnection studies [5]. There are still multiple modeling gaps that need to be addressed; if not, they can lead to inaccurate reliability studies, and consequently to unintentional load shedding or even blackouts. These gaps are mainly a consequence of decoupling hydro conditions and hydro-based constraints with electrical models used in steady-state and dynamic studies [6]. In this paper we will discuss the following gaps which have been identified in close collaboration with the industry and the utilities:

- Water availability in steady-state and dynamic models is not properly represented.
- Interdependences among hydro projects are not properly represented.
- Environmental constraints are not represented in models.
- Rough zones are not represented in the power system model so generation dispatch in system studies might not be realistic.
- Many outdated dynamic models of hydropower generation turbines are still in use.
- The power system is not studied for various water availability, such as high- and low-water conditions, in combination with a large amount of renewable generation.

To conduct more accurate planning and operation studies, and ensure the grid operates reliably, it is vital to model hydropower generation with more accuracy, particularly water availability. The intent is for steady-state and time-domain simulation of the system to match specific water conditions as closely as possible and avoid unrealistic expectations from hydro-based generation. This work addresses modeling gaps and provides recommendations for mitigation of the same.

This paper is structured as follows: in Section II we identify the modelling gaps in hydropower generation. Section III emphasizes the role of hydropower generation in supporting the power system with emphasize on Western Interconnection; and finally, in Section IV we provide recommendations for their mitigation and in Section V we conclude the paper with major findings and future work.

## II. Gaps in Hydropower generation Representation in Power System Models and Studies

While hydropower generation has been studied for decades, the dependency of power generation on water availability and constraints in hydro operation is not well illustrated in planning and operation models. In this section, we illustrate some of the gaps that need to be addressed. These gaps were identified through numerous discussions with industry partners and simulation studies.

### A. Water Availability (impact on $P_{max}$ and frequency)

The industry struggles to accurately model water availability and seasonal variations. Water level and availability depend on several factors such as season, amount of precipitation, snowpack, and temperature. Typically, the lowest water availability is during winter months when reservoirs are prepared for spring runoff. Hydropower generation output directly depends on the value of the water head. The head is the change in water levels between the water intake and the hydro discharge point. More water head means higher water pressure across the turbine so more power can be generated. Lower water levels in a reservoir means less generation will be available. In steady-state models, it is assumed the generator can provide nominal output, based on its capability curves, which is not always the case. As water levels vary annually and with seasons, the hydrogenator's available capacity needs to be adjusted to reflect water availability. These errors sum with each plant within an affected water basin.

Consequently, for low-water conditions, capacity and frequency response are overall very optimistic in operation and planning studies. Some plants have large water head variations. Figure 1 illustrates seasonal variations of the forebay for Grand Coulee Dam over a 10-year period. As the water head for plants having short penstock is calculated as forebay water minus tailwater, we can estimate that water head variation can be 70 feet. The nominal water head for Grand Coulee Dam is 330 feet, so a decrease of 70 feet is a 21% decrease in water head. As the maximum mechanical output of the plant is proportional to the power of 3/2 of the water head and function of water flow, for maximum flow through turbine $P_{max}$=f (1)$h^{1.5}$, plant output should be derated more than 30%. With 24 available units now, the maximum available power would be 4,545MW at peak flow.

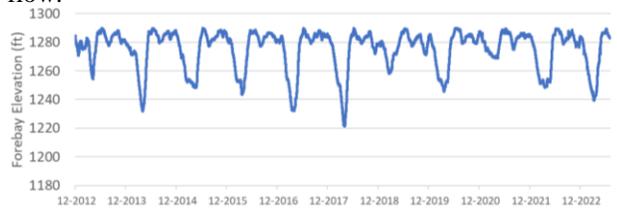

Figure 1. Forebay water variation for Grand Coulee Dam [7].

Similarly, water head values in dynamic models should be updated. As the generation is tested every 5 to 10 years, the same water head parameters stay in dynamic models for years. Power system planning studies are performed for heavy- and light-loading conditions, assuming nominal water availability. Some dynamic models do not allow changes in the water head, while others keep the head at 1 p.u. Consequently, frequency response might be overly optimistic. Environmental changes such as global warming further exacerbate the impact of water head. Variation in the water head affects the turbine operation and behavior, such discussions are not currently captured and lead to wrongful estimation of hydropower availability and planning as shown in Figure 2.

An extended drought and record heat wave in 2021 have pushed the water supply at Northern California's Lake Oroville to deplete to "alarming levels," which forced the closure of the Edward Hyatt Power Plant. The Colorado River basin is in a similar situation. Obviously, this water condition should be accounted for in planning and operational studies, since in this case power must be provided to load centers from other available sources, and this can lead to congestion and create reliability issues. The governor response becomes more of a concern with the increase in the renewable fleet and retirement of conventional generation. With low heads, the gate might reach maximum early in response, and MW response will be limited as a result.

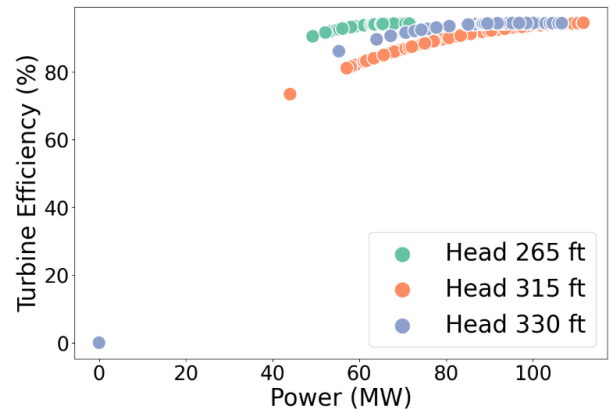

Figure 2. Francis's turbine efficiency under varying water head.

### B. Interdependences Among Cascading Plants

The Columbia River watershed provides more than 40% of the total U.S. hydropower generation. The Columbia River is highly regulated due to flood protection, transport needs, and federal laws. Most hydro projects are owned and operated by the U.S. Army Corps of Engineers and the Bureau of Reclamation. Management of cascading hydro projects is very complex and subject to many constraints based on the Endangered Species Act, which requires flow to support fish migration and spawning, and the Clean Water Act, which imposes minimum generation conditions. Water propagations need to consider (i.e., maintain the same ratio) between plant flows since many projects are on the river and do not have any or have very small reservoir capabilities. All these constraints are not represented in power flow. As shown in Figure 3, varying head during one season can create differences in generation pattern.

### A. Rough Zones

Generation operators are very familiar with rough zones; however, software tools used in operation and planning do not have information on rough zones for specific generators. The consequence of operating in a rough zone can be forced oscillation or damage to generation equipment. In general, generator operators tend to avoid operations in rough zones and try to run through as fast as possible. At some plants, there are automatic controls that detect rough operations and change setpoints until they smooth out. One of the more troublesome types of rough zones for both Francis and fixed-blade propeller turbines is the "*vortexing*" zone, a hydraulic phenomenon that can cause relatively large power swings (>5%) on generator output at low frequencies (one-quarter of the turbine shaft speed). This can occur on some turbines (mostly Francis) when operating in the mid-gate loading region like 40–60% wicket gate opening. As rough zones are not represented in the power-flow model, hydro-based generators can be dispatched inappropriately. Another issue is that models used in dynamic simulation cannot reproduce this effect of operating in rough zones.

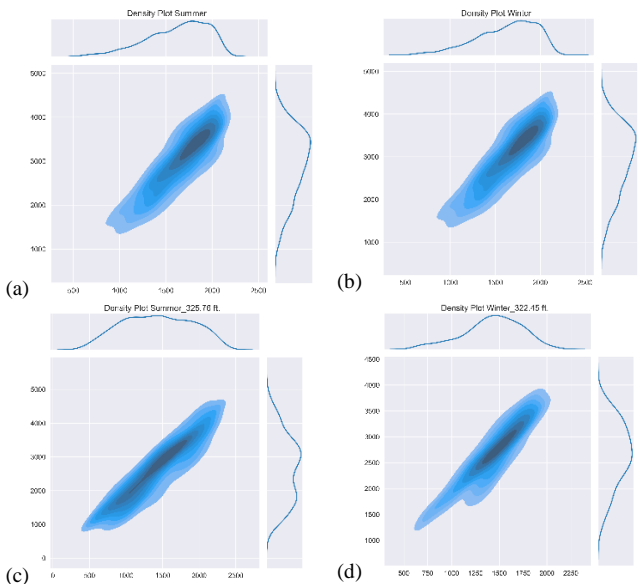

Figure 3. (a, b) Interdependency of two plants during different seasons. (c, d) Interdependency variation during a season with change in water head.

Operation in cavitation zones is also avoided, which typically happens at the more extreme upper and lower loading zones on a hydro machine and can cause more structural damage to the draft tube, head cover, penstock, and other large machine structures that support the weight of the turbine and rotor. As it is anticipated that hydropower generators are going to provide more regulation in the future system, it would be important to prevent generators from operating in these non-desired bands of operation. Operating in rough zones reduces turbine life expectancy in the long run, and interpreting different efficiency regions for turbine operations helps to refine its operation.

### B. Dynamic Models of Hydropower Generation

Many hydro-governor models are obsolete but are still used in practice. Some of them were developed in the late 1960s or early 1970s. For example, the GE PSLF and dynamic models *ieeeg3*, *pidgov*, *gpwscc*, and *g2wscc* do not have water head as a variable or they are a linearized model of turbine. As shown in Figure 2, operational efficiency and generated power vary greatly with head variation. As the turbine model is linearized, the model parameters are good only for small changes around given operating points. Their parameters are only for tested water conditions, so if the water conditions change (e.g., water head), they are not accurate representations of the turbine. Some more advanced models do not allow changes and keep the water head at 1 p.u. continuously.

### C. Inaccurate Representation of Operating Efficiencies

To calculate effective power output from hydropower turbines, the turbine efficiencies are assumed to be constant. The performance efficiency of the commonly used hydraulic turbines like *Propellor*, *Kaplan*, and *Francis* varies with respect to discharge profiles and water head values. Data collected for a site near Belknap Spring, Oregon, illustrated in Figure 4. It shows how the efficiency curves would vary for the same head and rated output for different turbine types. To assess their performance, design charts have been developed by turbine manufacturers to estimate efficiencies for rated flow and head. Interpreting such charts might be difficult for everyday use in power flow studies.

A data-driven approach to estimate the hydropower efficiency curves is discussed in [8], but the approach is limited to accessing high-quality data. A more generalized approach to estimating hydropower efficiency and power output is presented in [9]. It considers the complexities involving the different flow ranges and head to estimate a generalized or time series hydropower output. The time series estimate of power along with efficiency would be beneficial for running quasi-time series simulations. The approximation of efficiency ranges provides a better understanding of the different operational ranges of a turbine. This factor is essential, especially when hydro turbines are made to respond to fluctuating market signals. Multiple operations in certain zones might lead to early degradation of the turbines and increased maintenance and downtime.

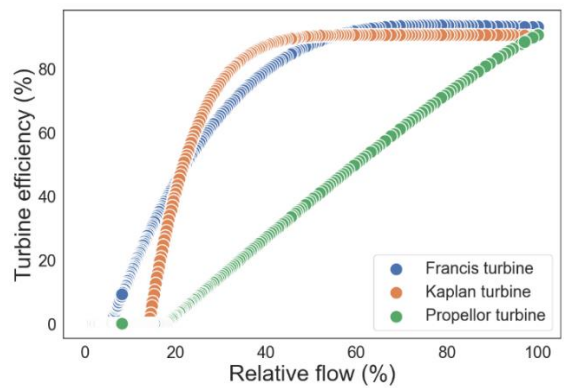

Figure 4. Variation in operational efficiency for different turbine types.

### III. DISTRIBUTION AND ROLE OF HYDROPOWER RESOURCES

About half of the total U.S. utility-scale conventional hydrogeneration capacity is concentrated in the states of Washington, California, and Oregon. Washington State produces the highest level of hydropower, contributing to 30 percent of the total hydropower in the United States and 68 percent of total electricity generation in the state. Similarly, Oregon, Idaho, Montana, and California contribute around 54, 56, 34, and 7 percent of total state production of hydroelectricity, respectively [10]. As such, the role of hydro in the Western Interconnection is crucial to keep lights on and provide base-loading and ancillary services. As the penetration of renewables increases, the role of hydropower becomes more stretched. In this section, we emphasize the role of hydropower in supporting the power system operations.

### A. Frequency Response

Intermittent renewable resources currently do not provide frequency response. In the future power system having solar and wind penetration frequency response becomes of more concern. Following system disturbance, such as the loss of a large generator, due to an imbalance of power, the frequency will decay. The power mismatch is initially compensated by extracting energy from rotational masses (inertial response) causing a decline in frequency. The decrease in frequency causes a decrease in load (load frequency dependency). If equilibrium is not reached and if frequency deviation is larger than governor dead-band settings, governors will respond by increasing the output of the units based on their droop characteristics. Frequency decay will be arrested when the equilibrium between generation and load is reached. This equilibrium is below the initial frequency. Automatic generation control will restore the frequency and interchange on scheduled value.

*1) Inertia*

System inertia affects the speed of frequency decline. The larger the inertia of the system, the slower the frequency of decay will be for the same amount of generation tripped. Hydropower generation is a significant source of inertia in the power system. The inertia constant $H$ of a typical hydro unit, when normalized per unit rating, is generally smaller than that of a thermal unit [11] due to the slower rotational speed of hydro turbines. However, as total inertia depends on the number of units online, knowing that a large hydro project can have more than a dozen generators compared to a typical steam-driven plant that typically has two to four units, the inertia of hydro projects can be significantly larger per project.

For large hydro plants with multiple units, the total stored energy in the rotational masses can be higher than in large nuclear/thermal facilities. With the replacement of conventional steam resources (thermal and nuclear) with solar and wind, the role of hydro in providing system inertia is even more crucial. In the future, to increase the inertia of the system, a hydropower generation owner might be encouraged to deploy more hydropower generators having lower output rather than deploying fewer generators having larger output.

*2) Primary response*

The primary frequency or governor response is the ability of the turbine generator set to measure changes in frequency and in the mechanical-power output of the turbine generator set and to adjust the mechanical power of the turbine based on generator droop characteristics. Fossil fuel plants typically burn coal or crude oil, and nuclear plants use nuclear energy to heat a boiler that produces high-temperature, high-pressure steam that is passed through the turbine to produce mechanical energy. The operation of steam turbines is challenging as the main steam temperature and pressure need to be maintained and governor response is provided by opening the main steam control valves. The process involves significant time delay, resulting in the governor action being slower. Most if not all the coal-fired and nuclear plants within the Western Interconnection have governors blocked so they do not provide primary frequency response. Gas turbines, on the other hand, can provide fast governor response and are considered an important resource to provide flexible generation and ramping capability. However, gas turbines are dependent on ambient temperature and their efficiency is affected by the temperature of the compressed air intake required to generate power.

For hydro turbines, the governor action results in changing the wicket gate position to change the flow of water, and consequently the real power output of the generator. Because hydro plant dynamics are non-minimum phase, power output will drop before rising when power is commanded to increase and vice versa.

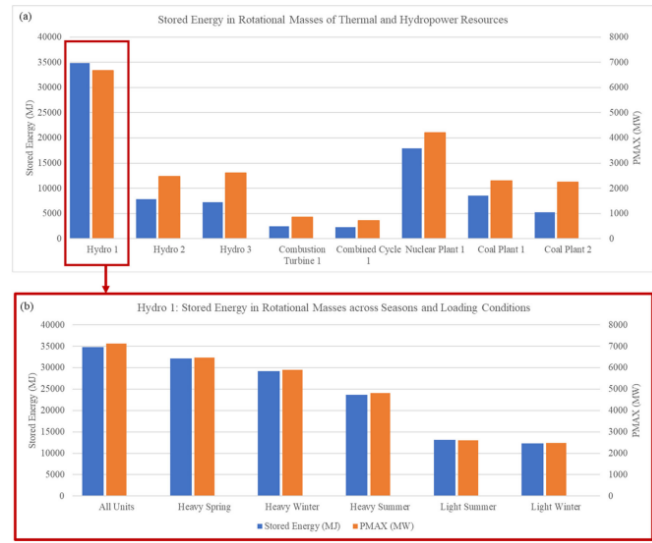

Figure 5. (a) Stored energy in rotational masses of thermal and hydropower resources in the Western Interconnection. (b) Hydro 1 hydropower resources available energy from inertia across seasons and loading. conditions.

Because of this, the hydro governor response is rate-limited to improve system stability, but after overcoming the initial response, the hydro ramping rate is faster. Hydropower generators are temperature-independent devices, making the efficiency of primary frequency response only affected by water availability. All installed hydro capacity cannot peak at the same time; nevertheless, hydropower generation from the Pacific Northwest typically provides a large share of primary frequency response as illustrated in Figure 5.

Figure 6 shows the surge in power flow from the Pacific Northwest to California on the California-Oregon Interface immediately after generation tripping in the South. Generation from the Pacific Northwest provided 400 MW of response for a loss of 680 MW in the South. It is worth mentioning that coal and nuclear power plants in the Western Interconnection do not respond and most of the Pacific Northwest generation fleet is hydro from the Columbia basin [12].

*3) Frequency ride-through and black-start capability*

One of the major concerns in power systems is that during large frequency disturbances, generators can trip due to under frequency protection of turbines. Under-frequency load shedding is supposed to prevent generator tripping, but if it is not fast enough—i.e., not enough load can be shed in time—or if the disturbance is too large, generators can trip due to under frequency protections leading to additional frequency drop and eventually to frequency collapse.

Hydropower generators have an advantage over other resources because they can operate over a significantly wider frequency range; hence, they are much less sensitive to changes in system frequency. The reason is that hydro turbines are rotating at a much slower speed compared to steam and gas turbines. Steam and gas turbines are very sensitive to variations in speed changes and are prone to permanent damage if they operate above or below rated speeds for some cumulative time over their lifecycle [13].

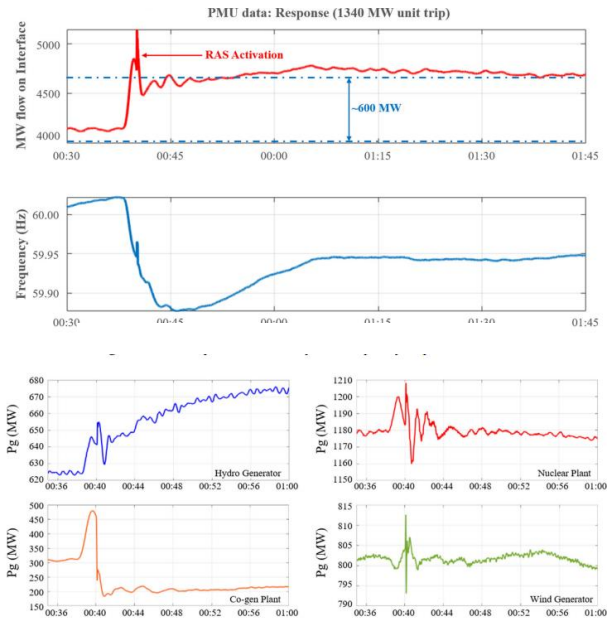

Figure 6. Hydropower's contribution toward primary frequency response.

On the other hand, hydropower turbines do not have such limitations. This advantage can be observed from the North American Electric Reliability Council (NERC) Protection and Control PRC-024 standard that provides frequency ride-through requirements for generators for different interconnections [14]. From generation frequency ride-through requirements for different interconnections the no-tripping zone ride-through requirements for Quebec were established to be much wider than the other interconnections (Western, Eastern, and the Electric Reliability Council of Texas) for both lower and higher frequency operation. The reason is that the Quebec interconnection is exclusively hydro and decoupled from the Eastern Interconnection through direct current links.

Because operating at low frequencies does not cause problems for hydro turbines, they do not have under-frequency protection. They might have only the quality protection that would reflect that, at some point, the performance of auxiliary loads would be affected, but only below the frequency levels at which under-frequency load shedding (steam turbine tripping) occurs. Loads like fans or pumps might have reduced flows due to lower frequency because these resources might eventually trip; however, these problems would not happen immediately, allowing additional time to ride through the event and allow the frequency to return to normal. During the 2003 blackout, when an estimated 50 million people were affected, hydropower facilities in northeastern states operated continuously and helped restore power [15].

## IV. Recommendations

Based on discussions with industry [1] and preliminary simulation studies, the most urgent need is getting ambient conditions and environmental constraints into the software used for power system operation and planning studies. This would make it possible to account for hydro conditions and constraints more accurately in studies; to specify season, month, day, and hour for a reasonable estimate of what generation capacity is available; and to make system studies more accurate, helping to increase the reliability of the system. To achieve this the following steps should be taken:

- Collect water data for different river basins for low- and high-water conditions.
- Collect rules on how water can be shifted from project to project.
- Collect environmental and other rules that are supported through water management.
- Develop tools that can impose water profile on each river basin separately and for individual hydro projects in steady-state and dynamic models for desired water conditions, that can implement coupling among plants in the same watershed and set operation points so that hydro units are not dispatched in rough zones [9].
- Replace classical linearized models with more appropriate models using changes in water head [16].

## V. Conclusions and Future Work

Reliability studies related to operation and planning do not currently incorporate environmental factors and limitations, such as the availability of water and the interdependencies between different power plants. This is because hydrological conditions and constraints are not effectively integrated with the electrical models utilized in power system planning and operational analyses, including power flow and dynamic models. The industry typically creates scenarios for extreme loading conditions during harsh winters and summers, as well as scenarios for lighter loading conditions. However, with the growing integration of intermittent renewable resources like solar and wind, along with the increasing emphasis on carbon-free objectives, hydropower's significance is heightened. Consequently, it's crucial to accurately account for hydropower conditions within the foundational scenarios. We are currently working on a tool that address the modelling gaps and incorporate different water dynamics, interdependency studies and other conditions while analyzing different contingency studies.